\documentclass[11pt,dvips,twoside,letterpaper]{article} 
\usepackage{times,fancyhdr}
\usepackage{graphicx}
\usepackage{geometry}
\usepackage{titlesec}

\titlelabel{\thetitle.\quad}                                

\usepackage[labelsep=period]{caption}                       


\makeatletter 
\def\cleardoublepage{\clearpage\if@twoside \ifodd\c@page\else%
    \hbox{}%
    \thispagestyle{empty}
    \newpage%
    \if@twocolumn\hbox{}\newpage\fi\fi\fi} 
\makeatother

\usepackage{amssymb}
\usepackage{amsmath}
\usepackage{amsfonts}
\usepackage{cite}
\usepackage{mathrsfs}

\renewcommand{\headrulewidth}{0pt}

\begin{document}
\title{
{\begin{flushleft}
\vskip 0.45in
{\normalsize\bfseries\textit{Chapter 1}}
\end{flushleft}
\vskip 0.45in
\bfseries\scshape Momentum-Field Interactions Beyond Standard Quadratic Optomechanics}}
\author{\bfseries\itshape Sina Khorasani\thanks{E-mail address: sina.khorasani@ieee.org.}\\
Vienna Center for Quantum Science and Technology, \\University of Vienna Boltzmanngasse, Vienna, Austria}
\date{}
\maketitle
\thispagestyle{empty}
\setcounter{page}{1}

\begin{abstract}
This chapter summarizes the recent progress in the theory and analytical tools of quadratic optomechanical interactions, as one of the prominent domains of contemporary nonlinear quantum optics. Emphasis has been put here first to show what types of nonlinear interactions do exist, and what physical interpretations follow each. The standard quadratic interactions between light and mechanical motion is expressed as the product of cavity light intensity and squared mirror position. However, there exists a non-standard quadratic optomechanical interaction as well, which assumes a mathematically different form and appears as the squared product of field and mirror momenta. This non-standard type of quadratic interaction originates from two corrections: the momentum exchange and conservation among mirror and field, as well as relativistic corrections due to different mechanisms. Both these types of non-standard interactions become relevant when the ratio of mechanical to optical frequency is no longer negligible. Next, we turn to the solution technique of such interactions, and introduce a formal higher-order operator method to tackle the nonlinear evolution of quantum systems. This enables one to accurately study any type of quantum nonlinear interaction using the analysis tools of linear algebra. In order to employ the analytical power of higher-order operator method, one first needs to identify a closed Lie algebra, which should satisfy closedness property under commutation either exactly or approximately, and is referred to as the basis. Having the basis of higher-order operators known, one may proceed to construct the corresponding Langevin equations, which can be now conveniently analyzed using the existing mathematical toolbox of linear algebra to yield the spectral densities, moments, and expectation values. Application of this operator method results in a new type of symmetry breaking in standard nonlinear quantum optomechanics, referred to as the sideband inequivalence.
\end{abstract}

\noindent {\bf PACS:} 42.50.-p, 42.50.Lc, 02.30.Tb

\vspace{.08in} 
\noindent {\bf Keywords:} quantum optomechanics, nonlinear interactions, Langevin equations


\pagestyle{fancy}
\fancyhead{}
\fancyhead[EC]{\it Sina Khorasani}
\fancyhead[EL,OR]{\thepage}
\fancyhead[OC]{\it Momentum-Field Interactions Beyond Standard Quadratic ...}
\fancyfoot{}
\renewcommand\headrulewidth{0pt} 


\section{Introduction}

Nonlinear interactions are at the heart of many physical phenomena, both in the classical and quantum regimes. It is the way physical quantities are described, that sets border between mathematical methods of classical and quantum physics. Classical quantities are described as real-valued scalars belonging to $\mathscr{R}$, while quantum quantities are described by operators belonging to an associated Hilbert space such as $\mathscr{C}$. Mathematically, any operator is equivalent to a finite- or infinite-dimensional matrix of complex scalar values. While a Fermionic operator such as an atomic transition and spin component operator, is normally finite-dimensional in its Heisenberg matrix representation, in which a bosonic operator such as ladder creator and annihilator becomes essentially infinite-dimensional. 

An interaction between two physical quantities $A$ and $B$ is contingent on the existence of algebraic expressions in the system Hamiltonian such as $f_{nm} A^n B^m$, where $n,m\in\mathscr{N}$ are natural numbers and $f_{nm}$ represents the interaction strength. For the interaction strength to make mathematical sense, it is necessary to be independent of $A$ and $B$. That implies $\partial f_{nm}/\partial A=\partial f_{nm}/\partial B=0$. There is furthermore no relation between $f_{nm}$ and $f_{mn}$ in general, and these two can be completely independent. 

When $n=m=1$, the interaction is referred to be linear, and otherwise nonlinear. The integer $o=n+m$ shall represent the interaction order, and in that sense, any nonlinear interaction demands a minimum order of $o\geq 3$. Any other type of nonlinear interaction not basically complying to the form $A^n B^m$ can be normally expanded using appropriate polynomial expansion methods as $\sum f_{nm} A^n B^m$ over non-zero integers $n,m\in\mathscr{Z}-\{0\}$ around some expansion point $(\bar{A},\bar{B})\in\mathscr{R}\times\mathscr{R}$. The interaction order needs then to be redefined as $|n|+|m|$.

We refer to the interaction among two quantities as bipartite. Definition of interaction between three and more physical quantities can be extended and generalized in a similar way, but a tripartite interaction such as $ABC$ can never be linear, unless decomposable into summation of bipartite linear interactions such as $f_{11}AB+g_{11}BC+h_{11}CA$ and so on. Solution of tripartite interactions is beyond the scope of present discussion, and we limit ourselves only to bipartite interactions which happen to be the case in optomechanics.

\subsection{Linear Interactions}

Nonlinear classical problems normally turn into the form of scalar nonlinear differential equations, which can be solved by conventional numerical methods. However, description of nonlinear quantum phenomena always turn into operator differential equations, which are basically a system of infinite-dimensional matrix equations wherever one interacting partition is a bosonic field with ladder operators involved. 

As opposed to the linear quantum interactions which involve direct products of two operator quantities such as $\hat{a}\hat{b}$ and its adjoint $\hat{b}^\dagger\hat{a}^\dagger$, nonlinear quantum interactions involve terms of at least third in the order such as $\hat{a}^2\hat{b}$ and the corresponding interaction Hamiltoninan is composed of a third-degree polynomial in terms of the most basic operators. Henceforth, a physical and Hermitian interaction having the type
\begin{equation}
\label{eq1}
\mathbb{H}_\text{int}=\hbar(\alpha\hat{a}\hat{b}+\beta\hat{a}^\dagger\hat{b}+\beta^\ast\hat{a}\hat{b}^\dagger+\alpha^\ast\hat{a}^\dagger\hat{b}^\dagger),
\end{equation}
where $\alpha,\beta\in\mathscr{C}$ are complex numbers, is having the most general form of linear interactions in quantum mechanics. It is easy to show that this interaction can be always factorized if $|\beta|=|\alpha|$. A factorizable bipartite interaction can be normally simplified using definition of linearly combined new operators, as $\hbar\xi\hat{x}\hat{y}$, where $\xi\in\mathscr{R}$ is a real number, and  $\hat{x}=\chi\hat{a}+\chi^\ast\hat{a}^\dagger$ and $\hat{y}=\eta\hat{b}+\eta^\ast\hat{b}^\dagger$ are Hermitian operators. Here, $\chi,\eta\in\mathscr{C}$ need to be half-modular complex constants satisfying $|\chi|^2=|\eta|^2=\frac{1}{2}$. This will set $\xi$ as the linear interaction strength.

\subsection{Quantum Optomechanics}

Sometimes, the interaction is nonlinear in terms of some quantities or operators, but assumes a linear form by defining some appropriate higher-order quantitites operators. Quantum optomechanics is one such example of interaction, with bipartite and third-order nonlinearity, in which there exist two bosonic partitions interacting nonlinearly. These partitions correspond to electromagnetic and mechanical bosonic fields respectively composed of photons and phonon, which get correlated nonlinearly. The optomechanical interaction Hamiltonian is here denoted as \cite{1,2,3,4,5,6,7,8,9}
\begin{equation}
\label{eq2}
\mathbb{H}_0=-\hbar g_0\hat{a}^\dagger\hat{a}(\hat{b}+\hat{b}^\dagger),
\end{equation}
\noindent
where $g_0\in\mathscr{R}^+$ is referred to as the single-photon interaction rate. At the lowest-order, the basic optomechanical interactions which describes the effect of radiation pressure on mirror's position are third in order. But with redefining the Hermitian operator $\hat{n}=\hat{a}^\dagger\hat{a}$ it could be seen that the optomechanical interaction is actually linear between $\hat{n}$ and $\hat{b}$. This fact can later be used to make the optomechanical Hamiltonian (\ref{eq2}) integrable, to be discussed in \S\ref{Fourth}

Normally, linearization of only optical field ladder operator $\hat{a}$ is sufficient to reduce the interaction degree to two, thus making the system integrable under linearized basic optomechanical Hamiltonian expressible as 
\begin{equation}
\label{eq3}
\mathbb{H}_\text{lin}=-\hbar g(\hat{a}+\hat{a}^\dagger)(\hat{b}+\hat{b}^\dagger),
\end{equation}
where $g\propto g_0$ is the enhanced interaction rate. Further linearization of $\hat{b}$ does not discard any physics of standard optomechanics, and only can make the analysis a bit simpler if needed. This already proves to be quite sufficient for clear understanding of many basic optomechanical pheonmena in the quantum regime. 

But further application of this linearization to quadratic and higher-order optomechanical interactions \cite{Paper1,Paper2}, leaves behind only a simple linear interaction quite similar to the linearized basic optomechanics (\ref{eq3}) though with a different interaction rate. This highlights the fact that any such linearization for quadratic, quartic and higher-order quantum interactions simply should not be done at high fluctuation amplitudes. This is because of the underlying physics of such interactions having the fourth-order and the above, which is completely wiped out through this linearization process. That is known to be the major obstacle in any large amplitude analysis of quadratic interactions, since some of the nonlinear behavior should be somehow kept within the governing remaining equations. 

The use of second-order operators in optomechanics can preserve some information, which is reminiscent of the essentially nonlinear optomechanical interaction \cite{Paper3}. This has been done and effects such as zero-point induced spring effect have been found. However, appropriate redefinition of third- and fourth-order operators \cite{Paper3} ultimately allows full integrability as will be demonstrated later in the text. Without this technique of higher-order operator algebra, the exact solvability of a quantum optomechanical problem using available conventional tools is out of question.

The analysis of quadratic interaction in quantum optomechanics \cite{Paper4,Paper5,Bruschi,Quad} gets more complicated by the fact that there exists two mathematically distinct and different interaction types. The first being referred to as the standard quadratic interaction is the light-position quadratic term expressed as 
\begin{equation}
\label{eq4}
\mathbb{H}_1=\hbar g_1\hat{a}^\dagger\hat{a}(\hat{b}+\hat{b}^\dagger)^2,
\end{equation}
while the non-standard quadratic interaction, results from either momentum exchange between light and mirror or relativistic corrections, and can be written as \cite{Paper1}
\begin{equation}
\label{eq5}
\mathbb{H}_2=-\hbar g_2(\hat{a}-\hat{a}^\dagger)^2(\hat{b}-\hat{b}^\dagger)^2.
\end{equation}
Here, the strengths of standard and non-standard quadratic interactions are actually related as
\begin{equation}
\label{eq6}
g_2=\frac{1}{4}\left(\frac{\pi^2}{3}+\frac{1}{4}\right)\left(\frac{\Omega}{\omega}\right)^2 g_1,
\end{equation}
where $\omega$ and $\Omega$ respectively represent the optical and mechanical resonant angular frequencies of the cavity. It has to be mentioned that (\ref{eq5}) could be rewritten in the equivalent form 
\begin{equation}
\label{eq7}
\mathbb{H}_2=-\hbar g_2(\hat{a}+\hat{a}^\dagger)^2(\hat{b}-\hat{b}^\dagger)^2,
\end{equation}
since an arbitrary $\frac{\pi}{2}$ phase shift in $\hat{a}$ interchanges (\ref{eq7}) and (\ref{eq5}), while leaving the standard quadratic (\ref{eq4}) and even the basic standard (\ref{eq2}) optomechanical interactions unaltered. 

Hence, the overall optomechanical interaction, in a system composed of an optomechanical cavity and a laser drive, and up to the quartic \cite{Paper1} order can be written after redefining the corresponding sign conventions as
{\small\begin{eqnarray}
\label{eq8}
\mathbb{H}&=&\mathbb{H}_\text{s}-\mathbb{H}_0+(\mathbb{H}_1-\mathbb{H}_2)-(\mathbb{H}_3+\mathbb{H}_4)+\mathbb{H}_\text{d}, \\ \nonumber
\mathbb{H}_\text{s}&=&\hbar\omega(\hat{a}^\dagger\hat{a}+\frac{1}{2})+\hbar\Omega(\hat{b}^\dagger\hat{b}+\frac{1}{2}),\\ \nonumber
\mathbb{H}_0&=&\hbar g_0\hat{a}^\dagger\hat{a}(\hat{b}+\hat{b}^\dagger), \\ \nonumber
\mathbb{H}_1&=&\hbar g_1\hat{a}^\dagger\hat{a}(\hat{b}+\hat{b}^\dagger)^2,\\ \nonumber
\mathbb{H}_2&=&\hbar g_2(\hat{a}+\hat{a}^\dagger)^2(\hat{b}-\hat{b}^\dagger)^2,\\ \nonumber
\mathbb{H}_3&=&\hbar g_3\hat{a}^\dagger\hat{a}(\hat{b}+\hat{b}^\dagger)^3,\\ \nonumber
\mathbb{H}_4&=&\hbar g_4(\hat{a}+\hat{a}^\dagger)^2\\ \nonumber
&&\times\left[(\hat{b}-\hat{b}^\dagger)^2(\hat{b}+\hat{b}^\dagger)+(\hat{b}+\hat{b}^\dagger)(\hat{b}-\hat{b}^\dagger)^2+(\hat{b}-\hat{b}^\dagger)(\hat{b}+\hat{b}^\dagger)(\hat{b}-\hat{b}^\dagger)\right],\\ \nonumber
\mathbb{H}_\text{d}&=&i\hbar(\alpha^\ast e^{i \tilde{\omega} t}\hat{a}-\alpha e^{-i \tilde{\omega} t}\hat{a}^\dagger).
\end{eqnarray}}
Here, $\tilde{\omega}$ is the frequency of drive, which is mostly set in resonance with the cavity at $\tilde{\omega}=\omega$. There could be multiple drives with different frequencies and amplitudes present, which simply add up to the number of drive terms in the Hamiltonian (\ref{eq8}). For instance, pumping the cavity with a second drive at frequency $\omega-\Omega$ on red-side band causes cooling. In $\mathbb{H}_\text{s}$ it is customary to drop the zero-point energy $E_\text{zp}=\frac{1}{2}\hbar\omega+\frac{1}{2}\hbar\Omega$ as it has no effect on the system behavior.

It is furthermore possible to write for an idealized one-dimensional cavity with parallel mirrors \cite{Paper1}
\begin{equation}
\label{eq9}
g_1=\frac{x_\text{zp}}{l}g_0,
\end{equation}
where $x_\text{zp}=\sqrt{\hbar/m\Omega}$ is the zero-point displacement of the cavity and $l$ is the cavity separation, with $m$ being the motion effective mass.
Also, the standard $g_3$ and non-standard $g_4$ quartic interaction rates for such an ideal cavity are connected as
\begin{equation}
\label{eq10}
g_4=\frac{1}{3\sqrt{2}}g_3\frac{1}{4}\left(\frac{\pi^2}{3}+\frac{1}{4}\right)\left(\frac{\Omega}{\omega}\right)^2=\frac{1}{3\sqrt{2}}\frac{g_2}{g_1}g_3=\frac{1}{3\sqrt{2}}\left(\frac{x_\text{zp}}{l}\right) g_2.
\end{equation} 
It is thus generally correct, that quadratic interactions need to be weaker than optomechanical interactions by a factor of $x_\text{zp}/l$. Similarly, quartic interactions are weaker than quadratic interactions by the same factor of $x_\text{zp}/l$. However, this argument does not necessarily hold for non-ideal cavities, where transverse geometry can influence $g_0$ independently. However, the relationships (\ref{eq9},\ref{eq10}) still give a sense of why quadratic and quartic interactions get progressively weak with the order increasing. 

While $g_0$ can be engineered to be made identically zero, $g_1$ does never vanish. Hence, quadratic effects always survive regardless of the existence of optomechanical interactions. Quartic effects also normally expected to vanish when $g_0=0$, and that implies the ideal condition for observation of quadratic effects is to design the optomechanical system in such a way that $g_0=0$. Since, neither of the lower- or higher-order interactions than the quadratic would effectively exist. This is actually not as difficult as it seems. An odd-profiled mechanical mode will have zero overlap with an even profiled incident optical field in optomechanics, or with the microwave field in superconducting electromechanics. Membrane-in-the-middle set up and tuning to the first odd-shaped mechanical frequency can theoretically establish the condition to achieve $g_0=0$, in a rather convenient way.

In practice, a bit of the optomechanical interaction $g_0$ may survive because of fabrication errors and that could be a source of inconvenience in quadratic measurements. Unless a practical tuning method to completely discard $g_0$ is available, it is safe to keep track of possible contributions from a small optomechanical term $\mathbb{H}_0$ in (\ref{eq8}) along with the quadratic terms $\mathbb{H}_1$ and $\mathbb{H}_2$.

For the reasons discussed in the above, quartic effects will be neglected from now on, and we will focus on two separate cases: Basic optomechanics with the quadratic and higher-order interactions dropped; this is discussed in \S\ref{Opto} Quadratic optomechanics with all other interactions dropped; this is discussed in \S\ref{Quad} Corrections to the quadratic interactions as a result of non-vanishing $g_0$ is discussed in \S\ref{OptoQuad}.

\subsection{Langevin Equations}

The analysis of a given Hamiltonian is done here through the well-known method of Langevin equations, which provide the equation of motion for operators in an open-system with input and output channels, interacting with a bath. This will cause a constant supply of fluctuating noise from each of the input channels to the system.

The corresponding Langevin \cite{Noise0,Noise1,Noise2,Noise3} equations to a given operator $\hat{z}$ of arbitrary order in an open system with multiple inputs are
\begin{eqnarray}
\label{eq11}
\dot{\hat{z}}&=&-\frac{i}{\hbar}[\hat{z},\mathbb{H}]\\ \nonumber
&+&\sum_j\left\{-[\hat{z},\hat{x}_j^\dagger]\left(\frac{\gamma_j}{2}\hat{x}_j+\sqrt{\gamma_j}\hat{x}_{j,\text{in}}\right)+\left(\frac{\gamma_j}{2}\hat{x}_j^\dagger+\sqrt{\gamma_j}\hat{x}_{j,\text{in}}^\dagger\right)[\hat{z},\hat{x}_j]\right\},
\end{eqnarray}
\noindent
where $\hat{x}_j$ is a system operator with the decay rate $\gamma_j$ and input flux $\hat{x}_{j,\text{in}}$. For optomechanical problems, the summation can be run over the bath operators $\hat{a}$ and $\hat{b}$. This method also enables one to construct the noise term of any higher-order operator $\hat{z}$ in a straightforward manner. This gives
\begin{equation}
\label{eq12}
\sqrt{\gamma}\hat{z}_\text{in}=\sum_j\sqrt{\gamma_j}\left\{-[\hat{z},\hat{x}_j^\dagger]\hat{x}_{j,\text{in}}+\hat{x}_{j,\text{in}}^\dagger[\hat{z},\hat{x}_j]\right\},
\end{equation}
in which $\gamma=\sum_j n_j\gamma_j$ is the effective decay rate of the higher-order operator $\hat{z}$ with $n_j\in\mathscr{N}$ being typically some natural numbers.

\section{Standard Optomechanics}\label{Opto}

In standard optomechanics, where no quardratic and higher-order terms exist, the Hamiltonian simply is
\begin{equation}
\label{eq13}
\mathbb{H}=\hbar\omega\hat{a}^\dagger\hat{a}+\hbar\Omega\hat{b}^\dagger\hat{b}-\hbar g_0\hat{a}^\dagger\hat{a}(\hat{b}+\hat{b}^\dagger)-\hbar(\alpha e^{i \tilde{\omega} t}\hat{a}+\alpha^\ast e^{-i \tilde{\omega} t}\hat{a}^\dagger).
\end{equation}
Here, the drive makes the Hamiltonian time-dependent at the frequency of $\tilde{\omega}$. For the usual case of resonant drive with $\tilde{\omega}=\omega$, the time-dependence of the Hamiltonian can be removed by transformation to the rotating frame as $\hat{a}\to e^{-i\omega t}\hat{a}(t)$, which makes the operator $\hat{a}(t)$ explicitly time-dependent, after removing a fast oscillation part. However, for the non-resonant drives also this is equally helpful, and nonetheless will not change the normal commutation relationship $[\hat{a}(t),\hat{a}^\dagger(t)]=1$. This will however change the time-derivative of the transformed operator $\hat{a}(t)$ as
\begin{equation}
\label{eq14}
\dot{\hat{a}}=\frac{d}{dt}\left[e^{-i\tilde{\omega} t}\hat{a}(t)\right]=e^{-i\tilde{\omega} t}\left[\dot{\hat{a}}(t)-i\tilde{\omega}\hat{a}(t)\right].
\end{equation}
While this transformation to the rotating frame is helpful with the standard interactions $\mathbb{H}_0$, $\mathbb{H}_1$, and $\mathbb{H}_3$, the algebraic form of non-standard interactions $\mathbb{H}_2$ and $\mathbb{H}_4$ excludes usefulness of this transformation. However, the non-standard interactions become only significant when $\omega$ is not far larger than $\Omega$, hence any further notion of fast oscillating part is point-less.

\subsection{Linear Optomechanics}

In stadard optomechanics, the Langevin equations have to be set up for the first-order basis $\{A\}^\text{T}=\{\hat{a},\hat{a}^\dagger,\hat{b},\hat{b}^\dagger\}$. This chosen basis is the lowest order possible of basis and the obvious choice for analysis of optomechanics. However, the Hamiltonian $\mathbb{H}_0$ is still nonlinear and the resulting Langevin equations will remain nonlinear. Calculation of (\ref{eq13}) for each of bath operators by setting $\hat{x}\in\{A\}$ gives four coupled equations. Henceforth, the exact Langevin equations while using (\ref{eq14}) are
\begin{eqnarray}
\label{eq15}
\dot{\hat{a}}&=&\left(i\Delta-\frac{1}{2}\kappa\right) \hat{a}+i g_0\hat{a}(\hat{b}+\hat{b}^\dagger)-\alpha-\sqrt{\kappa}\hat{a}_\text{in}, \\ \nonumber
\dot{\hat{a}}^\dagger&=&\left(-i\Delta-\frac{1}{2}\kappa\right) \hat{a}^\dagger-i g_0\hat{a}^\dagger(\hat{b}+\hat{b}^\dagger)-\alpha^\ast-\sqrt{\kappa}\hat{a}^\dagger_\text{in}, \\ \nonumber
\dot{\hat{b}}&=&\left(-i\Omega-\frac{1}{2}\Gamma\right) \hat{b}+i g_0\hat{a}^\dagger\hat{a}-\sqrt{\Gamma}\hat{b}_\text{in}, \\ \nonumber
\dot{\hat{b}}^\dagger&=&\left(i\Omega-\frac{1}{2}\Gamma\right) \hat{b}^\dagger-i g_0\hat{a}^\dagger\hat{a}-\sqrt{\Gamma}\hat{b}^\dagger_\text{in}.
\end{eqnarray}
where $\Delta=\tilde{\omega}-\omega$ is the optical detuning from drive frequency, $\kappa$ and $\Gamma$ are respectively the decay rates of $\hat{a}$ and $\hat{b}$, and the explicit time-dependence of operators after the transformation to the rotating frame has not been shown. 

The set of equations (\ref{eq14}) can be rewritten in the nonlinear matrix form
\begin{equation}
\frac{d}{dt}\{A\}=[\hat{\textbf{M}}]\{A\}-\sqrt{[\hat{\gamma}]}\{A_\text{in}\}-\{A_\text{d}\},
\end{equation}
Furthermore, $\{A_\text{d}\}^\text{T}=\{\alpha,\alpha^\ast,0,0\}$ is the drive input vector. It is easy to set up the elements of $[\hat{\textbf{M}}]$ by inspection from (\ref{eq15}), and there is of course no unique way to write it down because of the terms such as $\hat{a}\hat{b}$, whose either of their first or second operators could be put within the coefficients matrix. 

The resulting Langevin equations for the first-order basis $\{A\}$ is nonlinear because of the presence of second-order operators $\hat{a}\hat{b}$ and $\hat{a}\hat{b}^\dagger$ in the first equation, $\hat{a}^\dagger\hat{b}$ and $\hat{a}^\dagger\hat{b}^\dagger$ in the second equation, and $\hat{a}^\dagger\hat{a}$ in the last two equations. Therefore, (\ref{eq15}) is not integrable, unless the nonlinear terms can be appropriately simplified.

The first step to linearize (\ref{eq15}) is to do the replacement $\hat{a}\to\bar{a}+\hat{a}$, where $\bar{a}$ is the average value of the field operator. We here may suppose that  $\bar{a}\in\mathscr{R}$ is taken to be real-valued, as its corresponding phase can be adjusted in the pump $\alpha$. Then, we can linearize the photon number $\hat{a}^\dagger\hat{a}\to\bar{a}^2+\bar{a}(\hat{a}+\hat{a}^\dagger)$, which drives the radiation pressure term with the time average $\bar{n}=\bar{a}^2$. The presence of an average radiation pressure term, puts a constant displacement upon mirror $\bar{b}$ around which the mirror movements fluctuate. Hence, we arrive at a similar linearization of $\hat{b}\to\bar{b}+\hat{b}$ and thus $\hat{a}\hat{b}\to\bar{a}\hat{b}+\bar{b}\hat{a}$.

Once these replacements are plugged in (\ref{eq15}), the Langevin equations are linearized and the static expressions in terms of average values can be separated to yield
\begin{eqnarray}
\label{eq16}
\left(i\Delta-\frac{1}{2}\kappa\right) \bar{a}+i 2g_0\Re[\bar{b}]\bar{a}=\alpha, \\ \nonumber\left(-i\Omega-\frac{1}{2}\Gamma\right) \bar{b}=-i g_0\bar{n}.
\end{eqnarray}
These equations can be combined to obtain the real-valued $\bar{n}=\bar{a}^2$ in terms of $|\alpha|$ through solution of a third-order algebraic equation, given as
\begin{equation}
\label{eq17}
\left(\Delta+ \frac{4g_0^2\Omega}{\Omega^2+\frac{1}{4}\Gamma^2}\bar{n}\right)^2\bar{n} +\frac{\kappa^2}{4}\bar{n}=|\alpha|^2.
\end{equation}
The linearized $4\times 4$ Langevin equations around the average values now take the form
\begin{eqnarray}
\label{eq18}
\dot{\hat{a}}&=&\left[i(\Delta+f)-\frac{1}{2}\kappa\right] \hat{a}+i g(\hat{b}+\hat{b}^\dagger)-\sqrt{\kappa}\hat{a}_\text{in}, \\ \nonumber
\dot{\hat{a}}^\dagger&=&\left[-i(\Delta+f)-\frac{1}{2}\kappa\right] \hat{a}^\dagger-i g(\hat{b}+\hat{b}^\dagger)-\sqrt{\kappa}\hat{a}^\dagger_\text{in}, \\ \nonumber
\dot{\hat{b}}&=&\left(-i\Omega-\frac{1}{2}\Gamma\right) \hat{b}+i g(\hat{a}+\hat{a}^\dagger)-\sqrt{\Gamma}\hat{b}_\text{in}, \\ \nonumber
\dot{\hat{b}}^\dagger&=&\left(i\Omega-\frac{1}{2}\Gamma\right) \hat{b}^\dagger-i g(\hat{a}+\hat{a}^\dagger)-\sqrt{\Gamma}\hat{b}^\dagger_\text{in}.
\end{eqnarray}
in which $g=g_0\bar{a}$ and $f=2g_0\Re[\bar{b}]=2g_0^2\Omega\bar{n}/(\Omega^2+\frac{1}{4}\Gamma^2)$. The analysis of the system of equations (\ref{eq18}) is extensively discussed elsewhere  \cite{1,2,3,4,6}. It is convenient, nevertheless, to put the integrable system (\ref{eq18}) in the generic matrix form and rewrite the linearized Langevin equation as
\begin{equation}
\label{eq19}
\frac{d}{dt}\{A\}=[\textbf{M}]\{A\}-\sqrt{[\gamma]}\{A_\text{in}\},
\end{equation}
with definitions
{\footnotesize\begin{eqnarray}
\label{eq20}
[\textbf{M}]&=&\left[
\begin{array}{cccc}
i(\Delta+f)-\frac{1}{2}\kappa & 0 & ig & ig \\ 
0 & -i(\Delta+f)-\frac{1}{2}\kappa & -ig & -ig \\
ig & ig & -(i\Omega+\frac{1}{2}\Gamma) & 0 \\
-ig & -ig & 0 & i\Omega-\frac{1}{2}\Gamma 
\end{array}
\right], \nonumber \\ \nonumber
[\gamma]&=&\text{Diag}\{\kappa,\kappa,\Gamma,\Gamma\}, \\ 
\{A_\text{in}\}^\text{T}&=&\{\hat{a}_\text{in},\hat{a}_\text{in}^\dagger,\hat{b}_\text{in},\hat{b}_\text{in}^\dagger \}.
\end{eqnarray}}
Now, input-output relations \cite{Noise1,Noise2} can be used with the definition of the scattering matrices in Fourier domain with probe detuning $w=\omega_\text{p}-\omega$ to yield
\begin{eqnarray}
\label{eq21}
\{A_\text{out}(w)\}&=&\{A_\text{in}(w)\}-\sqrt{[\gamma]}\{A(w)\}, \\ \nonumber
\{A_\text{out}(w)\}&=&[\textbf{S}(w)]\{A_\text{in}(w)\}, \\ \nonumber
[\textbf{S}(w)]&=&[\textbf{I}]-\sqrt{[\gamma]}\left(iw[\textbf{I}]-[\textbf{M}]\right)^{-1}\sqrt{[\gamma]}.
\end{eqnarray}
It is to be kept in mind that it is mostly the spectral density of $\hat{a}$ denoted by $\mathcal{S}_{AA}(w)$ which can be measured in an either heterodyne or homodyne experiment. However, the spectral density of input is known, which enables one to obtain the spectral density of output in terms of the input using the relation
\begin{equation}
\label{eq22}
\mathcal{S}_{AA}(w)=\sum_{j=1}^{4}|S_{1j}(w)|^2 \mathcal{S}_{j,\text{in}}(w),
\end{equation}
where $S_{1j}(w)$ are elements of the first row of the scattering matrix $[\textbf{S}(w)]$ and 
\begin{eqnarray}
\label{eq23}
\{\mathcal{S}_{\text{in}}(w>0)\}^\text{T}&=&\{1,0,\bar{m}+1,\bar{m}\}, \\ \nonumber
\{\mathcal{S}_{\text{in}}(w<0)\}^\text{T}&=&\{0,1,\bar{m},\bar{m}+1\},
\end{eqnarray}
is the spectral density vector of noise channel due to photonic and phononic baths. We here notice that $\mathscr{F}\{\hat{a}(t)\}(w)=\mathscr{F}\{\hat{a}^\dagger(t)\}^\ast(-w)$, and so on. Also, $\bar{m}$ represents the cavity phonon number or coherent phonon population, which is a different quantity than the thermal occupation number of phonons and corresponds to the phonons driven coherently by the optomechanical interaction.

\subsection{Second-Order Optomechanics}
The second-order equations of optomechanics are obtained by using the mixed first- and second-order basis $\{A\}^\text{T}=\{\hat{a},\hat{b},\hat{a}\hat{b},\hat{a}\hat{b}^\dagger,\hat{n},\hat{c}\}$ \cite{Paper3}, in which we have the definition $\hat{c}=\frac{1}{2}\hat{a}^2$ \cite{Paper1,Paper2}. It is easy to observe that this set forms a closed basis, since the commutator of every pair of operators belonging to this basis results from a linear combination of the operators therein. The non-zero commutators related to this basis are
\begin{eqnarray}
\label{eq24}
[\hat{c},\hat{n}]&=&2\hat{c}, \\ \nonumber
[\hat{a},\hat{n}]&=&\hat{a}, \\ \nonumber
[\hat{b},\hat{a}\hat{b}^\dagger]&=&\hat{a},\\ \nonumber
[\hat{a}\hat{b},\hat{n}]&=&\hat{a}\hat{b},\\ \nonumber
[\hat{a}\hat{b}^\dagger,\hat{n}]&=&\hat{a}\hat{b}^\dagger,\\ \nonumber
[\hat{a}\hat{b},\hat{a}\hat{b}^\dagger]&=&2\hat{c}.
\end{eqnarray}
The resulting $6\times 6$ Langevin equations should be constructed based on the Hamiltonian (\ref{eq13}) and this has been extensively discussed elsewhere \cite{Paper3}. 

While these are nonlinearly coupled, further linearization of the system shall decouple three equations out of six as it will be shown shortly, leaving the optomechanical system expressible in terms of the reduced basis $\{A\}^\text{T}=\{\hat{a},\hat{a}\hat{b},\hat{a}\hat{b}^\dagger\}$, which is now not closed under commutation. Putting into the standard form (\ref{eq19}) and before linearization, the $3\times 3$ coefficients operator matrix $[\hat{\textbf{M}}]$ for the set of Langevin equations of the reduced basis looks like
{\small \begin{equation}
\label{eq25}
[\hat{\textbf{M}}]=
\left[ 
\begin{array}{ccc}
i\Delta-\frac{\kappa }{2} & ig_0 & ig_0 \\ 
ig_0\left(\hat{m}+\hat{n}+1\right)  & -i\left(\Omega-\Delta-g_0 \hat{b}\right)-\frac{\gamma }{2} & 0  \\ 
ig_0\left(\hat{m}-\hat{n}\right) &  0 & i\left(\Omega+\Delta+g_0\hat{b}^\dagger\right)-\frac{\gamma }{2} 
\end{array}
\right],
\end{equation}}
where $\gamma=\kappa+\Gamma$, with the decay rate matrix, and input noise and drive vectors
\begin{eqnarray}
\label{eq26}
[\hat{\gamma}]&=&\left[
\begin{array}{ccc}
\kappa & 0 & 0\\
\kappa\hat{b} & \Gamma\hat{a} & 0\\
\kappa\hat{b} & 0 & \Gamma\hat{a}
\end{array}
\right],\\ \nonumber
\{A_\text{in}\}^\text{T}&=&\{\hat{a}_\text{in},\hat{b}_\text{in},\hat{b}^\dagger_\text{in}\}, \\ \nonumber
\{A_\text{d}\}^\text{T}&=&\{\alpha,\alpha\hat{b},\alpha\hat{b}^\dagger\}.
\end{eqnarray}

After linearization of (\ref{eq25}) and (\ref{eq26}), it is possible to show that the steady state values of $\bar{b}$ and $\bar{a}=\sqrt{\bar{n}}$ must exactly satisfy (\ref{eq16}) and (\ref{eq17}) again. The first Langevin equation is already linear. However, linearization of second and third equations at this stage gives
\begin{eqnarray}
\label{eq27}
[\textbf{M}]&=&
\left[ 
\begin{array}{ccc}
i\Delta-\frac{\kappa }{2} & ig_0 & ig_0 \\ 
ig_0\left(\bar{m}+\bar{n}+1\right)  & -i\left(\Omega-\Delta\right)-\frac{\gamma }{2} & 0  \\ 
ig_0\left(\bar{m}-\bar{n}\right) &  0 & i\left(\Omega+\Delta\right)-\frac{\gamma }{2}  
\end{array}
\right],\\ \nonumber
[\gamma]&=&\left[
\begin{array}{ccc}
\kappa & 0 & 0\\
\kappa|\bar{b}| & \Gamma\sqrt{\bar{n}} & 0\\
\kappa|\bar{b}| & 0 & \Gamma\sqrt{\bar{n}}
\end{array}
\right], \\ \nonumber
\{A_\text{d}\}^\text{T}&=&\{\alpha,0,0\}.
\end{eqnarray}

\subsection{Fourth-Order Optomechanics}\label{Fourth}

It is possible to select a third-order closed basis of operators \cite{Paper3}, which can surprisingly lead to a fully linearized system of equations. The particular choice of $\{A\}^\text{T}=\{\hat{n}^2,\hat{n}\hat{b},\hat{n}\hat{b}^\dagger\}$ here is composed of a fourth-order operator $\hat{N}=\hat{n}^2$ and a third-order operator $\hat{B}=\hat{n}\hat{b}$ and its conjugate $\hat{B}^\dagger$. This system will lead to a linearized system of Langevin equations which is actually decoupled for $\hat{B}$ and $\hat{B}^\dagger$. This particular choice, referred to as the minimal basis, reduces to a simple $2\times 2$ Langevin system (\ref{eq15}) with
\begin{eqnarray}
\label{eq29}
\{A\}^\text{T}&=&\{\hat{N},\hat{B}\}\\ \nonumber
[\textbf{M}]&=&\left[
\begin{array}{cc}
-2\kappa & 0 \\
ig_0 & -i\Omega-\frac{\gamma}{2}
\end{array}
\right], \\ \nonumber
[\gamma]&=&\text{Diag}\{2\kappa,\gamma\}, \\ \nonumber
\{A_\text{in}\}^\text{T}&=&\{\hat{N}_\text{in},\hat{B}_\text{in}\} = \{\hat{n}\hat{a}^\dagger\hat{a}_\text{in}+\hat{a}_\text{in}^\dagger\hat{a}\hat{n},\sqrt{2\kappa}\hat{b}\hat{n}_\text{in}+\sqrt{\Gamma}\hat{n}\hat{b}_\text{in}\},\\ \nonumber
\{A_\text{d}\}^\text{T}&=&\{\alpha\hat{n}\hat{a}^\dagger+\alpha^\ast\hat{a}\hat{n},\alpha\hat{b}\hat{a}^\dagger+\alpha^\ast\hat{b}\hat{a}\}.
\end{eqnarray}
It is now fairly easy and quite straightforward to construct the \textit{exact and explicit solution} to the system (\ref{eq29}) as
\begin{eqnarray}
\label{eq30}
\hat{N}(t)&=&\hat{N}(0)e^{-2\kappa t}-2\sqrt{\kappa}\int_{0}^{t}e^{-2\kappa (t-\tau)}\hat{N}_\text{in}(\tau)d\tau, \\ \nonumber
\hat{B}(t)&=&\hat{B}(0)e^{-(i\Omega+\frac{\gamma}{2}) t}-\int_{0}^{t}e^{-(i\Omega+\frac{\gamma}{2}) (t-\tau)}\left[ig_0\hat{N}(\tau)+\sqrt{\gamma}\hat{B}_\text{in}(\tau)\right]d\tau.
\end{eqnarray}
The multiplicative noise terms \cite{Noise4} and drive vector in (\ref{eq29}) can be approximated as
\begin{eqnarray}
\label{eq31}
\{A_\text{in}\}^\text{T}&=&\left\{\frac{\sqrt{\bar{n}}\bar{n}}{2}\left(\hat{a}_\text{in}+\hat{a}_\text{in}^\dagger\right),\sqrt{\frac{\kappa}{\gamma}\bar{n}}\bar{b}\left(\hat{a}_\text{in}+\hat{a}_\text{in}^\dagger\right)+\sqrt{\frac{\Gamma}{\gamma}}\bar{n}\hat{b}_\text{in}\right\}, \\ \nonumber
\{A_\text{d}\}^\text{T}&=&2\sqrt{\bar{n}}\{\bar{n},\bar{b}\}\Re[\alpha].
\end{eqnarray}

\section{Quadratic Optomechanics}\label{Quad}

Solution of quadratic optomechanics needs a different basis of higher-order operators \cite{Paper2}. The basis with minimum dimension required to deal with the Hamiltonain (\ref{eq8}) is \cite{Paper4}
\begin{equation}
\label{eq32}
\{A\}^\text{T}=\{\hat{c},\hat{c}^\dagger,\hat{n},\hat{d},\hat{d}^\dagger,\hat{m}\},
\end{equation}
in which $\hat{d}=\frac{1}{2}\hat{b}^2$ and $\hat{m}=\hat{b}^\dagger\hat{b}$. This basis is also closed under commutation and satisfies a Lie algebra. To see this, it is sufficient to verify the commutators $[\hat{c},\hat{n}]=2\hat{c}$ from (\ref{eq24}), which gives $[\hat{n},\hat{c}^\dagger]=2\hat{c}^\dagger$, and $[\hat{c},\hat{c}^\dagger]=\hat{n}+\frac{1}{2}$ for photons, and in a similar manner $[\hat{d},\hat{d}^\dagger]=\hat{m}+\frac{1}{2}$,
$[\hat{d},\hat{m}]=2\hat{d}$, and $[\hat{m},\hat{d}^\dagger]=2\hat{d}^\dagger$ for phonons. All other commutators within the above basis are zero.

First, the coefficients of Langevin equations (\ref{eq16}) can be partitioned as
\begin{equation}
\label{eq33}
[\hat{\textbf{M}}]=\left[
\begin{array}{c|c}
\hat{\textbf{M}}_\text{aa} & \hat{\textbf{M}}_\text{ab} \\ \hline
\hat{\textbf{M}}_\text{ba} & \hat{\textbf{M}}_\text{bb}
\end{array}\right],
\end{equation}
with indices a and b referring to photons and phonons. The $3\times 3$ partitions are now given by 
{\small\begin{eqnarray}
\label{eq34}
[\hat{\textbf{M}}_\text{aa}]&=&\left[
\begin{array}{ccc}
-2i\omega-\kappa & 0 & i\frac{1}{2} g_2(\hat{f}-\hat{m})\\
0 & 2i\omega-\kappa & -i\frac{1}{2} g_2(\hat{f}-\hat{m})\\
-ig_2(\hat{f}-\hat{m}) & ig_2(\hat{f}-\hat{m}) & -\kappa
\end{array}
\right] , \\ \nonumber
[\hat{\textbf{M}}_\text{ab}]&=&ig_2 \left[
\begin{array}{ccc}
-\beta_-(\hat{c}-\hat{c}^\dagger) & -\beta_-(\hat{c}-\hat{c}^\dagger) & \beta_+(\hat{c}-\hat{c}^\dagger) \\
\beta_-(\hat{c}-\hat{c}^\dagger) & \beta_-(\hat{c}-\hat{c}^\dagger) & -\beta_+(\hat{c}-\hat{c}^\dagger) \\
0 & 0 & 0
\end{array}
\right], \\ \nonumber
[\hat{\textbf{M}}_\text{ba}]&=&\frac{ig_2}{2}\left[
\begin{array}{ccc}
\hat{m}-2\hat{d}+\frac{1}{2} & \hat{m}-2\hat{d}+\frac{1}{2} & \hat{m}-2\hat{d}+\frac{1}{2} \\
-\hat{m}+2\hat{d}^\dagger-\frac{1}{2} & -\hat{m}+2\hat{d}^\dagger-\frac{1}{2} & -\hat{m}+2\hat{d}^\dagger-\frac{1}{2} \\
0 & 0 & 0
\end{array}
\right], \\ \nonumber
[\hat{\textbf{M}}_\text{bb}]&=&\left[
\begin{array}{ccc}
-2i\Omega-\Gamma & 0 & -i\frac{1}{2}g_2\hat{n}\\
0 & 2i\Omega-\Gamma & i\frac{1}{2}g_2\hat{n}\\
-ig_2(\hat{e}-\beta_-\hat{n})& ig_2(\hat{e}-\beta_-\hat{n}) & -\Gamma
\end{array}
\right].
\end{eqnarray}}
Here, we have adopted the notions  $\beta_\pm=(g_1/g_2)\pm 1$ as well as $\hat{e}=\hat{c}+\hat{c}^\dagger$  and $\hat{f}=\hat{d}+\hat{d}^\dagger$. It is to be noticed that there is no unique way to partition the above nonlinear system. Meanwhile, the decay matrix, multiplicative input noise, and drive vector are
\begin{eqnarray}
\label{eq35}
[\gamma]&=&\text{Diag}\{\kappa,\kappa,\kappa,\Gamma,\Gamma,\Gamma \}, \\ \nonumber
\{A_\text{in}\}^\text{T}&=&\{\hat{a}\hat{a}_\text{in},\hat{a}_\text{in}^\dagger\hat{a}^\dagger,\hat{a}^\dagger\hat{a}_\text{in}+\hat{a}_\text{in}^\dagger\hat{a},\hat{b}\hat{b}_\text{in},\hat{b}_\text{in}^\dagger\hat{b}^\dagger,\hat{b}^\dagger\hat{b}_\text{in}+\hat{b}_\text{in}^\dagger\hat{b} \}, \\ \nonumber
\{A_\text{d}\}^\text{T}&=&\{i\alpha e^{-i\omega t}\hat{a},i\alpha^\ast e^{i\omega t}\hat{a}^\dagger, \alpha e^{-i\omega t}\hat{a}+\alpha^\ast e^{i\omega t}\hat{a}^\dagger,0,0,0\}.
\end{eqnarray}

We also here have to notice that the particular form of quadratic Hamiltonian $\mathbb{H}_2$ excludes any usefulness of transformation to rotating frame of coordinate, and the optical frequency here is measured absolutely. 

Linearization of the corresponding Langevin equations around the mean values gives rise to the partitions \cite{Paper4}

{\small\begin{eqnarray}
\label{eq36}
[\textbf{M}_\text{aa}]&=&\left[
\begin{array}{ccc}
i2(\omega-g_2\beta_+\bar{m})-\kappa & 0 & -i\frac{1}{2}g_2\bar{m} \\ 
0 & -i2(\omega-g_2\beta_+\bar{m})-\kappa & i\frac{1}{2}g_2\bar{m} \\ 
ig_2\bar{m} & -ig_2\bar{m} & -\kappa 
\end{array}
\right] , \\ \nonumber
[\textbf{M}_\text{ab}]&=&i\frac{g_2}{2}\left(\bar{n}+\frac{1}{2}\right)\left[
\begin{array}{ccc}
1 & 1 & -1 \\ 
-1 & -1 & 1 \\ 
0 & 0 & 0 \end{array}
\right], \\ \nonumber
[\textbf{M}_\text{ba}]&=&ig_2\left(\bar{m}+\frac{1}{2}\right)\left[
\begin{array}{ccc}
1 & 1 & -\beta_- \\
-1 & -1 & \beta_- \\
0 & 0 & 0 
\end{array}
\right], \\ \nonumber
[\textbf{M}_\text{bb}]&=&\left[
\begin{array}{ccc}
-i2(\Omega+g_2\beta_+\bar{n})-\Gamma & 0 & -ig_2\beta_-\bar{n} \\
0 & i2(\Omega+g_2\beta_+\bar{n})-\Gamma & ig_2\beta_-\bar{n} \\
i2g_2\beta_-\bar{n} & -i2g_2\beta_-\bar{n} & -\Gamma
\end{array}
\right].
\end{eqnarray}}
Here, photon and phonon frequencies already have shifted to the new values $\omega+g_1-2g_2\to\omega$, and $\Omega-2g_2\to\Omega$, and we have assumed that the optical drive is actually resonant with the newly shifted value. The input vector and decay matrix have to be reformatted as \cite{Paper4}
\begin{eqnarray}
\label{eq37}
\{A_\text{in}\}^\text{T}&=&\left\{
\hat{a}_\text{in},
\hat{a}_\text{in}^\dagger, 
\frac{1}{2}(\hat{a}_\text{in}+\hat{a}_\text{in}^\dagger), 
\hat{b}_\text{in},
\hat{b}_\text{in}^\dagger,
\frac{1}{2}(\hat{b}_\text{in}+\hat{b}_\text{in}^\dagger) 
\right\}, \\ \nonumber
[\gamma]&=&\text{Diag}\left[\bar{n}\kappa,\bar{n}\kappa,4\bar{n}\kappa,2|\bar{d}|\Gamma,2|\bar{d}|\Gamma,4|\bar{d}|\Gamma\right].
\end{eqnarray}
Finally, the mean photon and phonon values are to be found from the nonlinearly coupled algebraic equations \cite{Paper4}
\begin{eqnarray}
\label{eq38}
4|\alpha|^2&=&(g_1+g_2)^2\bar{m}^2\bar{n}, \\ \nonumber
4\bar{n}|\alpha|^2&=&g_1^2(\bar{m}^2-\bar{m}),
\end{eqnarray}
together with $4|\bar{d}|^2=\bar{m}^2-\bar{m}$, required to complete the set of equations (\ref{eq37}). Combining these two equations leads for the near resonant case to the fourth-order algebraic equation in terms of $\sqrt{\bar{n}}$ as 
\begin{equation}
\label{eq37a}
2|\alpha|\bar{n}^2+\frac{g_1^2}{g_1+g_2}\sqrt{\bar{n}}=g_1^2\frac{2|\alpha|}{(g_1+g_2)^2}.
\end{equation}
This equation reveals that in the limit of very strong near-resonant pumping, the cavity photon population gets saturated to the value $\bar{n}_\text{max}=(1+\rho)^{-1}$ where $\rho=g_2/g_1$. This particular behavior of cavity photon number has already been observed under numerical simulations \cite{Paper4}. However, cavity phonon number $\bar{m}$ can still continue to increase almost proportional to the input photon flux as $\bar{m}\propto|\alpha|$. If the pump frequency is not exactly tuned to the shifted cavity frequency as described under (\ref{eq36}), therefore we always have $\bar{n}_\text{max}<1$, with no apparent limit on $\bar{m}$.  In absence of momentum interactions with $g_2=0$, the cavity photon number will be clamped to $\bar{n}=1$ via non-resonant quadratic interactions. 

If the pump is however accurately tuned to the altered cavity optical frequency $\omega+g_1-2g_2$ to obtain the on-resonance criteria, it has been shown \cite{Paper4} that achieving this condition can result in a virtually unlimited increase in steady-state cavity photon population $\bar{n}$ while phonon population  gets saturated around unity with $\bar{m}\approx 1$.

This surprising behavior at resonance and off-resonance or near-resonance reveals that an observable quadratic effect could be expected even if the cavity is not pumped resonantly, since anyhow the cross population $\psi=\bar{m}\bar{n}$ tends to increase with the pump $|\alpha|$. If tuned to the shifted resonance, this product increases as $\psi\propto|\alpha|^2$. 

This mysterious behavior can explain the fact that quadratic interactions do not need resonant pumping, and once they are strong enough and the pumping level is sufficiently high, they should be observable. This fact has also been demonstrated in the numerical studies of this phenomenon \cite{Paper4}.

One may therefore take the cross population $\psi$ as a measure of strength of quadratic interactions. At high pump levels we have for an on-resonance relationship $\psi\propto|\alpha|^2$, while the off-resonance relationship reads $\psi\propto|\alpha|$.

\section{Perturbations from Basic Optomechanics}\label{OptoQuad}

It is useful from a practical point of view to know to what extent and how a quadratic optomechanical interaction $\mathbb{H}_1-\mathbb{H}_2$ could be influenced and even possibly masked by a non-vanishing basic optomechanical interaction $\mathbb{H}_0$. This will enable one to rigorously differentiate between the type and strength of standard basic optomechanical and quadratic optomechanical responses to an excitation. Since the non-vanishing $g_0$ is expected to be randomly non-zero from one fabricated device to the other, the possible range of non-zero $g_0$ could be scanned and the responses should be averaged to come up with a practically meaningful expectation for the possible response.

In order for this to be done, one would need to include the optomechanical interaction $\mathbb{H}_0$ in the formulation as well. Proceeding with the same basis as (\ref{eq32}) is possible but with the minor correction to the coefficients matrix (\ref{eq34}) and (\ref{eq36}) as well as the equations for steady-state populations (\ref{eq38}). 

It is straightforward to calculate the change in each of the four partitions as follows before and after linearization. The exact change in (\ref{eq34}) before linearization is given by
\begin{eqnarray}
\label{eq39}
[\Delta\hat{\textbf{M}}_\text{aa}]&=&2ig_0\left[
\begin{array}{ccc}
\hat{b}+\hat{b}^\dagger & 0 & 0 \\
0 & -\hat{b}-\hat{b}^\dagger & 0 \\
0 & 0 & 0
\end{array}
\right], \\ \nonumber
[\Delta\hat{\textbf{M}}_\text{ba}]&=&ig_0\left[
\begin{array}{ccc}
0 & 0 & \hat{b} \\
0 & 0 & -\hat{b}^\dagger \\
0 & 0 & -\hat{b}+\hat{b}^\dagger
\end{array}
\right], 
\end{eqnarray}
while $[\Delta\hat{\textbf{M}}_\text{ab}]=[\Delta\hat{\textbf{M}}_\text{bb}]=[\textbf{0}]$. It is seen that $[\Delta\hat{\textbf{M}}]$ is unsurprisingly proportional to the non-vanishing optomechanical interaction rate $g_0$. These relations can be further linearized as 
\begin{eqnarray}
\label{eq40}
[\Delta\hat{\textbf{M}}_\text{aa}]&=&4i\Re[\bar{b}]g_0\left[
\begin{array}{ccc}
1 & 0 & 0 \\
0 & -1 & 0 \\
0 & 0 & 0
\end{array}
\right], \\ \nonumber
[\Delta\hat{\textbf{M}}_\text{ba}]&=&ig_0\left[
\begin{array}{ccc}
0 & 0 & \bar{b} \\
0 & 0 & -\bar{b} \\
0 & 0 & -2i\Im[\bar{b}]
\end{array}
\right].
\end{eqnarray}
Here, $\bar{b}$ shall be determined from the second relation of (\ref{eq17}) as
\begin{equation}
\label{eq41}
\bar{b}=\frac{ig_0\bar{n}}{i\Omega+\frac{1}{2}\Gamma},
\end{equation}
together with (\ref{eq38}) for $\bar{n}$, $\bar{m}$, and $\bar{d}$. It should be noticed that in the ideal mathematical sense, one would expect $\bar{b}=\sqrt{\bar{d}}$. However, for a nearly vanishing $g_0$ with indeterminate and uncertain value close to zero, this approximation should be enough to give a sense of a basis optomechanical interaction entering into the picture. The reason is simply that any actual error in this calculation of $\bar{b}$ could be directly attributed to magnitude $g_0$. 

When pumping strongly on resonance with $g_0$ not vanishing, the optomechanical tri-state equation (\ref{eq17}) and the quadratic state equation (\ref{eq38}) compete to determine the photon population $\bar{n}$. One from (\ref{eq17}) may expect that $\bar{n}\propto\sqrt[3]{|\alpha|^2}$, while it can be shown by inspection of governing state equation \cite{Paper4} that for resonantly pumped quadratic interaction requires $\bar{n}\propto|\alpha|^{2}$ which grows more rapidly than the other effect. Surprisingly, quadratic interaction on resonance does not heat up the cavity, since phonon population would anyhow remains locked to a value very close to unity $\bar{m}\approx 1$. This behavior is similar to the weakly coupled optomechanics with $g=g_0\sqrt{\bar{n}}<<\sqrt{\Omega|\Delta|}$, which infers the same type of behavior according to (\ref{eq17}).

Hence, for a non-vanishing optomechanical interaction rate $g_0$, we can expect that at a sufficiently high pump level, and given that it is exactly tuned to the shifted cavity resonance, the optomechanical interaction will be perfectly masked by the quadratic interaction and therefore could be neglected and completely ignored. 

\subsection{Sideband Inequivalence}

As a remarkable conclusions which could be drawn from higher-order analysis of optomechanical systems is the inequivalence of red- and blue-sidebands. While it is a well-known fact that the amplitudes of these sidebands are a bit different due to finite thermal occupation number of phonons, it is almost unnoticed that the corresponding frequency shifts are also not exactly equal. This fact has been termed as sideband inequivalence and discussed in much details elsewhere \cite{Paper3}. As a simple approximation, the amount of non-zero sideband inequivalence can be estimated as $\delta\Omega\approx g_0^2\bar{n}/\Omega$, where $\delta\Omega=\frac{1}{2}(\Delta_r+\Delta_b)$ represents the non-zero average of red-shifted $\Delta_r$ and blue-shifted $\Delta_b$ detunings.

Obviously, this effect makes practical sense only for weakly-coupled and sideband resolved cavities, where $\Omega>>g_0\sqrt{\bar{n}}$ and $\Omega>>\kappa$ respectively hold. While one would expect that a large sideband inequivalence could be observed by increased intracavity photon number and/or pumping rate, this is practically not the case for Doppler cavities with $\kappa>\Omega$, where very large intracavity photon numbers $\bar{n}$ are attainable. 

Basically, a cavity which is not sideband resolved disallows clear observation of sidebands, as they are strongly suppressed by fluctuating noise. On the one hand, numerical tests reveal that sideband inequivalence does actually exist and should be mathematically strong in Doppler cavities. On the other hand, sidebands are strongly damped and suppressed, which implies that they cannot be resolved and probably are physically unimportant altogether. This combination of opposing facts shows that sideband resolved cavities must be used.

A sideband resolved cavity also sets a stringent limit on the maximum practically attainable intracavity photon number $\bar{n}$, and thus the expected amount of inequivalence $\delta\Omega$. However, the available enhanced interaction rate $G=g_0 \bar{n}$ is still limited, so that strongly coupled cavities with large $g_0$ can only accommodate very few intracavity photons $\bar{n}$. Again, there is a maximum practical limit on how large a practically observable sideband inequivalence could be. For all practical purposes, the normalized sideband inequivalence $\delta\Omega/\Omega$  for various types of optomechanical and electromechanical systems seem to be bounded to $10^{-6}$ to $10^{-4}$. 

\section*{Conclusion}

We described different approaches to solve the standard basic, and standard quadratic, and non-standard quadratic optomechanical problems. The conventional method of linearizing Langevin equations in terms of ladder operators may not be useful for study of quadratic and higher-order effects, however, definition of higher-order alternative bases could provide a means to conveniently analyze and understand higher-order nonlinear effects in quantum physics. It was shown that the cross population of photons and phonons could be taken as a measure of the strength of quadratic interactions, which increases porportional to or quadratically with respectively off-resonance and on-resonance strong pumping. We also argued that in the presence of a non-vanishing basic optomechanical interaction, on-resonance quadratic pumping could easily mask out the basic optomechanical interaction. More importantly, while the intensity of quadratic interactions increase with the square of input pump power at resonance, the cavity phonon number gets saturated normally to a sub-unity value. This highlights the fact that it is possible to observe quadratic interactions at high illumination power, without heating the cavity up.

\section*{Acknowledgments}

Discussions with Georg Arnold at Institute of Science and Technology Austria, Dr. David Edward Brushi at Vienna Center for Quantum Science and Technology, and Dr. Andr\'{e} Xuereb at University of Malta is appreciated. This work is dedicated to the celebrated artist, Anastasia Huppmann.

\bigskip


\label{lastpage-01}

\end{document}